\documentclass[3p,times,procedia]{elsarticle}
\usepackage{nupha_ecrc}

\volume{00}

\firstpage{1}

\journalname{Nuclear Physics A}

\runauth{Takahito Todoroki (for the STAR Collaboration)}

\jid{nupha}

\jnltitlelogo{Nuclear Physics A}

\usepackage{amssymb}

 \usepackage{lineno}

\usepackage[figuresright]{rotating}

\begin{document}

\begin{frontmatter}



\dochead{XXVIth International Conference on Ultrarelativistic Nucleus-Nucleus Collisions\\ (Quark Matter 2017)}

\title{
Measurements of charmonium production in p$+$p, p$+$Au, and Au$+$Au collisions at $\sqrt{s_{_{\rm NN}}}$~=~200~GeV with the STAR experiment
}


\author{Takahito Todoroki (for the STAR Collaboration) 
}

\address{Physics Department, Brookhaven National Laboratory, Upton, New York 11973, USA}

\begin{abstract}
We present the first results from the STAR MTD of mid-rapidity charmonium measurements via the di-muon decay channel in p$+$p, p$+$Au, and Au$+$Au collisions at $\sqrt{s_{_{\rm NN}}}=200$~GeV at RHIC.
The inclusive $J/\psi$ production cross section in p$+$p collisions can be described by
the Non-Relativistic QCD (NRQCD) formalism coupled with the color glass condensate effective theory (CGC) at low transverse momentum ($p_T$)
and next-to-leading order NRQCD at high $p_T$.
The nuclear modification factor in p$+$Au collisions for inclusive $J/\psi$
is below unity at low $p_T$ and consistent with unity at high $p_T$, which can be described by calculations including both nuclear PDF and nuclear absorption effects.
The double ratio of inclusive $J/\psi$ and $\psi(2S)$ production rates for $0<p_T<10$~GeV/$c$ at mid-rapidity between p$+$p and p$+$Au collisions is measured
to be 1.37~$\pm$~0.42~$\pm$~0.19.
The nuclear modification factor in Au$+$Au collisions for inclusive $J/\psi$ shows significant $J/\psi$ suppression at high $p_T$ in central collisions 
and can be qualitatively described by
transport models including dissociation and regeneration contributions.

\end{abstract}


\end{frontmatter}


\section{Introduction }
The $J/\psi$ dissociation by the color-screening effect in the hot and dense medium~\cite{Matsui:1986dk} was initially proposed as direct evidence of 
the quark-gluon plasma formation. However, the interpretation of $J/\psi$ suppression observed in heavy-ion collisions has remained a challenge due to the contribution of regenerated $J/\psi$
from the coalescence of deconfined $c\bar{c}$ pairs in the medium as well as cold nuclear matter effects.
Quantifying the cold and hot nuclear matter effects at the RHIC requires precise measurements of charmonium production
in p$+$p, p$+$Au, and Au$+$Au collisions.
The Muon Telescope Detector (MTD), which provides both 
the muon triggering and identification capabilities
at mid-rapidity,
opens the door to measuring quarkonia via the di-muon decay channel at STAR.
Using the MTD di-muon trigger, the STAR experiment recorded data corresponding to an integrated luminosity of
14.2~nb$^{-1}$ in Au$+$Au collisions at $\sqrt{s_{_{\rm NN}}}=200$~GeV in the RHIC 2014 run,
and integrated luminosities of 122~pb$^{-1}$ in p$+$p collisions
and 409~nb$^{-1}$ in p$+$Au collisions at $\sqrt{s_{_{\rm NN}}}=200$~GeV in the RHIC 2015 run.
In these proceedings, we present (i) measurements of nuclear modification factors for inclusive $J/\psi$ production over a broad kinematic range in
both p$+$Au and Au$+$Au collisions at $\sqrt{s_{_{\rm NN}}}=200$~GeV; and (ii) the first measurement of the double ratio of inclusive $\psi(2S)$ and $J/\psi$
production rates at mid-rapidity between p$+$p and p$+$Au collisions at $\sqrt{s_{_{\rm NN}}}=200$~GeV.

\section{Inclusive $J/\psi$ measurements in p$+$p and p$+$Au collisions at $\sqrt{s_{_{\rm NN}}}=200$~GeV }
\label{}
\begin{figure}[h]
\begin{minipage}{17pc}
\center
\includegraphics[width=13pc]{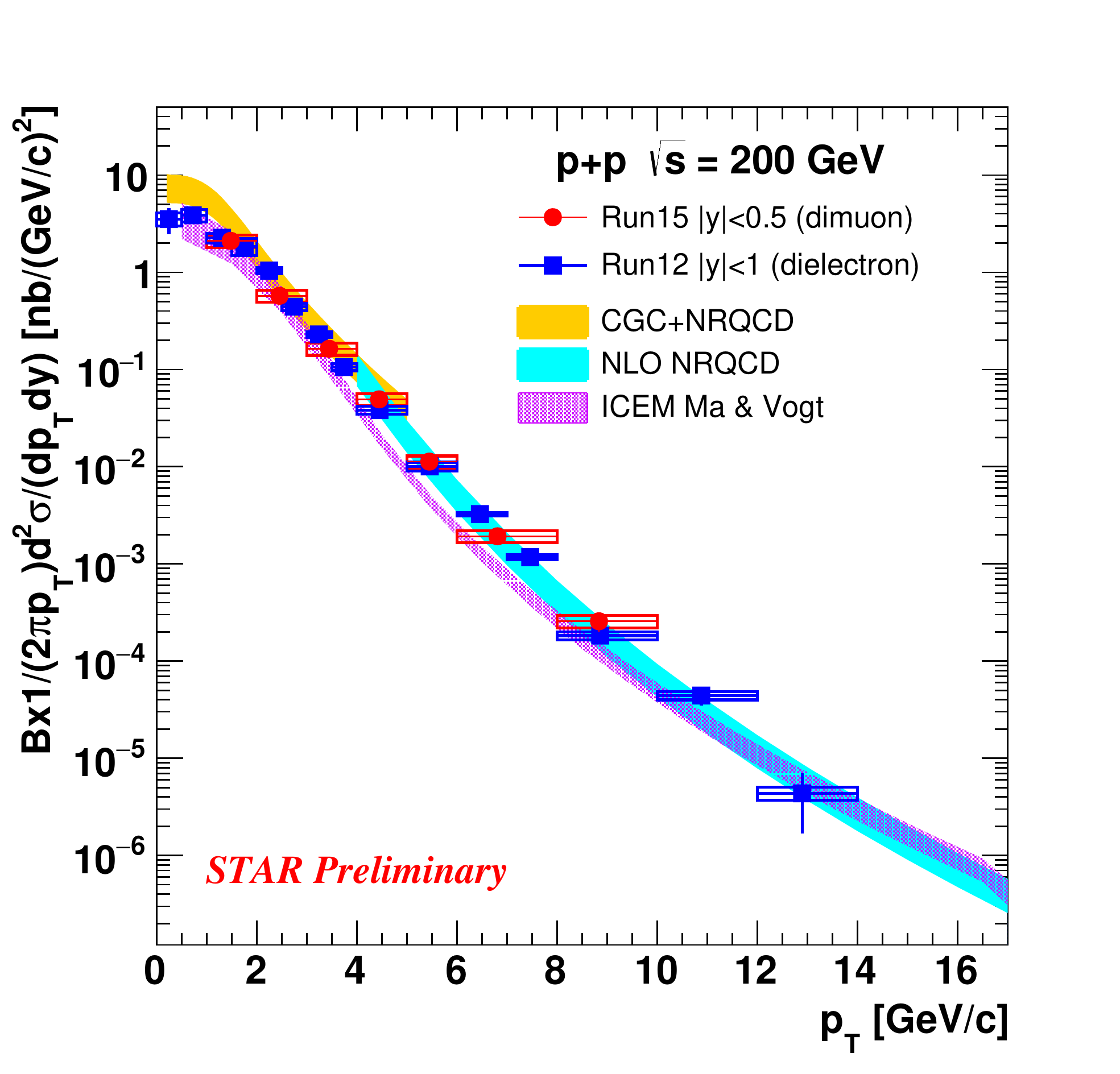}
\caption{\label{fig:Run15JpsiCrossSection} Inclusive $J/\psi$ production cross section scaled by the branching ratio ($B$) as a function of $p_T$ in the di-muon (red circle) 
and the di-electron decay channels (blue square).}
\end{minipage}\hspace{1.5pc}%
\begin{minipage}{17pc}
\center
\includegraphics[width=18pc]{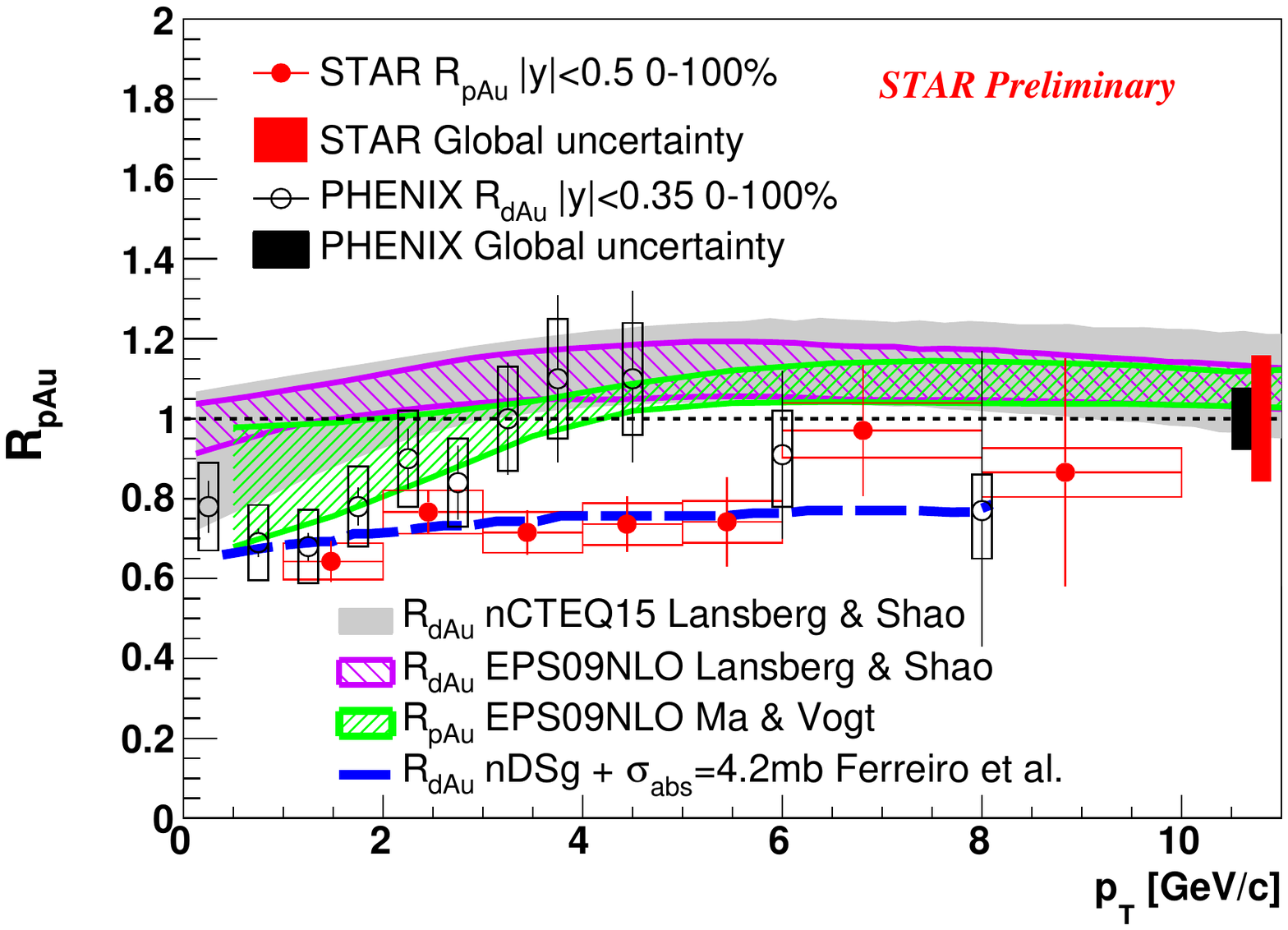}
\caption{\label{fig:Run15RpAu} Nuclear modification factor $R_{\rm pAu}$ as a function of $p_T$ for inclusive $J/\psi$ in the di-muon
 decay channel.}
\end{minipage} 
\end{figure}

Figure~\ref{fig:Run15JpsiCrossSection} shows the production cross section of inclusive $J/\psi$ in p$+$p collisions at $\sqrt{s}=200$~GeV
via the di-muon decay channel for the transverse momentum ($p_T$) range of $1<p_T<10$~GeV/c (red circles),
along with a similar measurement via the di-electron decay channel (blue squares) in $0<p_T<14$~GeV/c. 
These results are consistent in the overlapping $p_T$ range.
The experimental results can be well described by CGC+NRQCD~\cite{Ma:2014mri} and NLO NRQCD~\cite{Shao:2014yta} calculations for prompt $J/\psi$
at low and high $p_T$ ranges, respectively.
While an improved color evaporation model (ICEM) calculation for direct $J/\psi$~\cite{Ma:2016exq}
can describe the data for $p_T<3$~GeV/$c$, it generally underestimates the yield at higher $p_T$.

Figure~\ref{fig:Run15RpAu} shows the nuclear modification factor, $R_{\rm pAu}$,
of inclusive $J/\psi$ in 0-100\% central p$+$Au collisions. The measured $R_{\rm pAu}$ is generally consistent with the previous $R_{\rm dAu}$ result reported
by the PHENIX experiment~\cite{Adare:2012qf} within statistical and systematic uncertainties.
The largest deviation between these results is 1.4$\sigma$ in the range of $3<p_T<5$~GeV/$c$.
This overall consistency suggests similar cold nuclear matter effects in p$+$Au and d$+$Au collisions.
Calculations, taking into account the nuclear PDF effect using the nCTEQ15~\cite{Lansberg:2016deg,Shao:2012iz,Shao:2015vga}
or EPS09NLO~\cite{Lansberg:2016deg,Shao:2012iz,Shao:2015vga,Shao:2016private} nuclear PDF sets, can touch the upper limit of the data within uncertainties.
However, the model calculation including an additional nuclear absorption effect~\cite{Ferreiro:2012sy} is favored by the data.

\section{Double ratio of inclusive $J/\psi$ and $\psi(2S)$ yields between p$+$p and p$+$Au collisions at $\sqrt{s_{_{\rm NN}}}=200$~GeV}
\begin{figure}[h]
\begin{minipage}{17pc}
\center
\includegraphics[width=15pc]{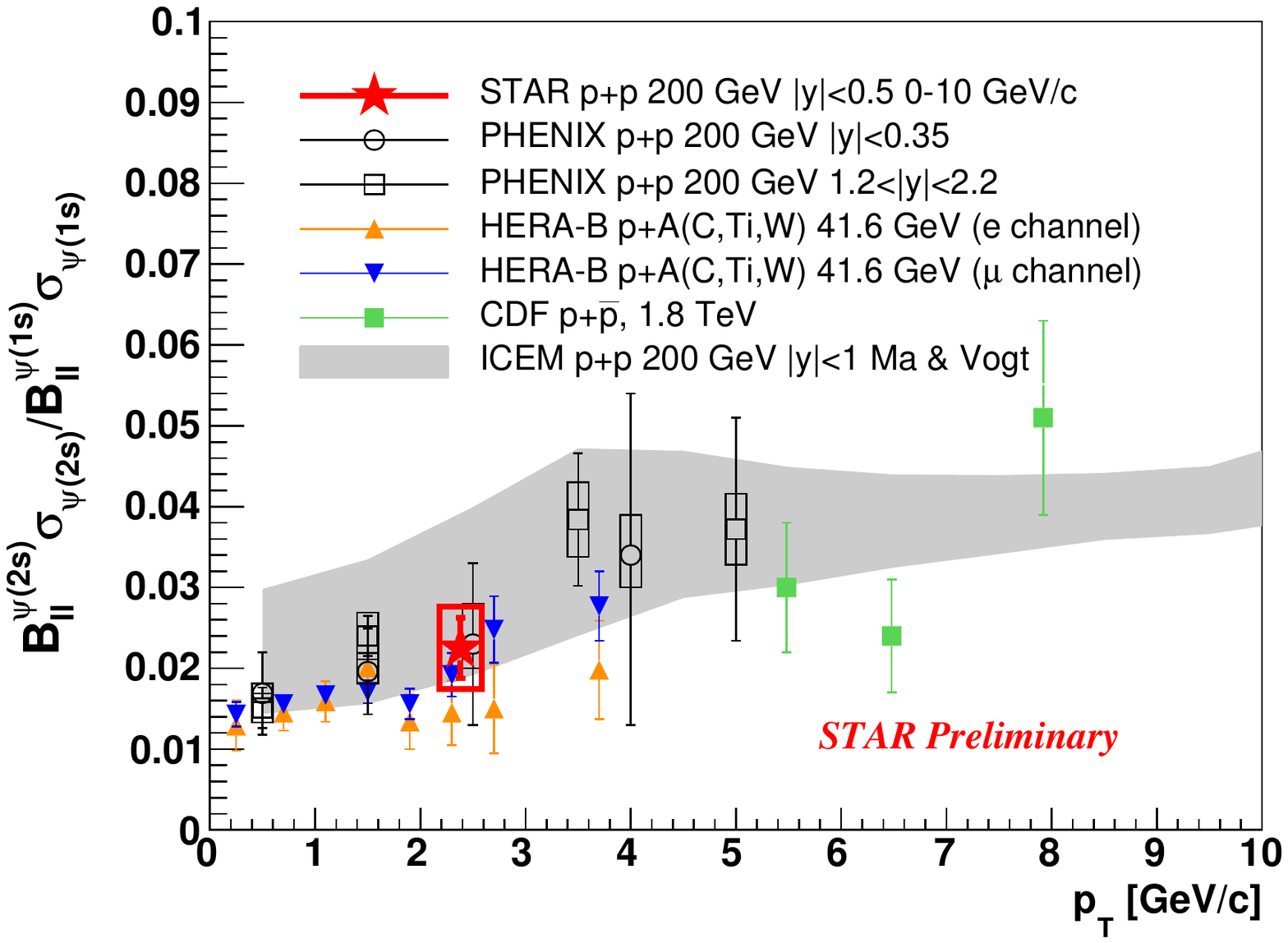}
\caption{\label{fig:Run15ppJpsiPsi2sRatio} 
Ratio of $\psi(2S)$ to $J/\psi$ production rates as a function of $p_T$.}
\end{minipage}\hspace{1.5pc}%
\begin{minipage}{17pc}
\center
\includegraphics[width=15pc]{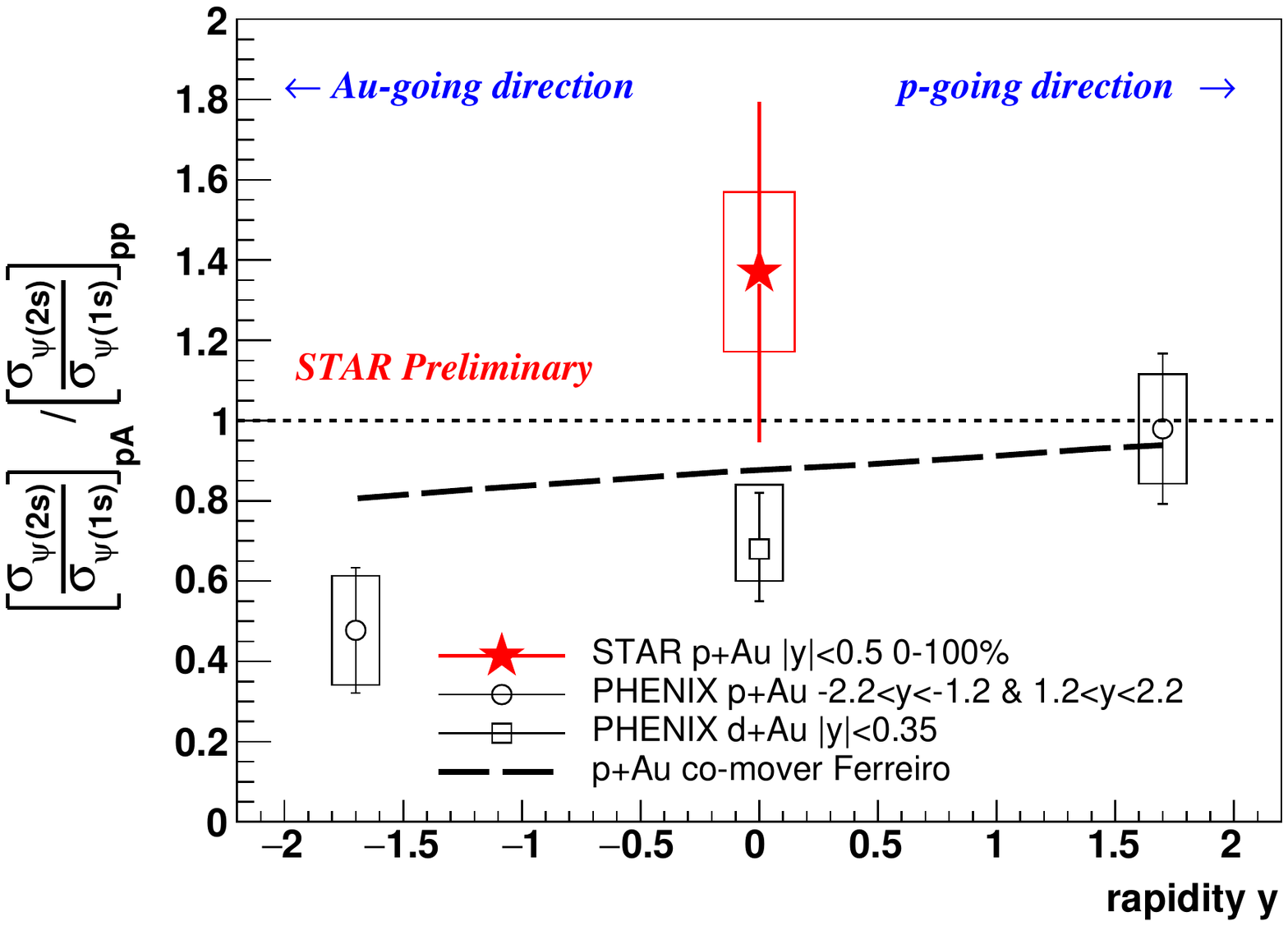}
\caption{\label{fig:Run15pppAuJpsiPsi2sDoublaRatio}
Double ratio of $J/\psi$ and $\psi(2S)$ production rates as a function of rapidity in p$+$Au and d$+$Au collisions.
}
\end{minipage} 
\end{figure}

Figure~\ref{fig:Run15ppJpsiPsi2sRatio} shows the ratio of inclusive $J/\psi$ and $\psi(2S)$ production cross sections 
as a function of $p_T$ in p$+$p collisions at $\sqrt{s}=200$~GeV.
The new STAR result for $0<p_T<10$~GeV/$c$ follows the global trend of results by HERA~\cite{Abt:2006va},
PHENIX~\cite{Adare:2011vq,Adare:2016psx}, and CDF~\cite{Abe:1997jz} experiments. The ICEM calculation at $\sqrt{s}=200$~GeV~\cite{Ma:2016exq} can describe 
the increasing trend of the ratio with $p_T$.

Figure~\ref{fig:Run15pppAuJpsiPsi2sDoublaRatio} shows the double ratio of $\psi(2S)$ and $J/\psi$ production rates
between p$+$p and p$+$Au collisions as a function of rapidity. The new STAR results at $|y|<0.5$ is 1.37~$\pm$~0.42(stat)~$\pm$~0.19(sys),
which is consistent with the published PHENIX results at $|y|<0.35$ in d$+$Au collisions~\cite{Adare:2013ezl}.
The co-mover model calculation~\cite{Ferreiro:2016private,Ferreiro:2014bia} can qualitatively
describe the double ratio at forward and backward rapidities 
in p$+$Au collisions reported by the PHENIX experiment~\cite{Adare:2016psx},
and is consistent with the new STAR result at mid-rapidity within uncertainties.

\section{Inclusive $J/\psi$ measurements in Au$+$Au collisions at $\sqrt{s_{_{\rm NN}}}=200$~GeV }
\begin{figure}[h]
\begin{minipage}{17pc}
\center
\includegraphics[width=15pc]{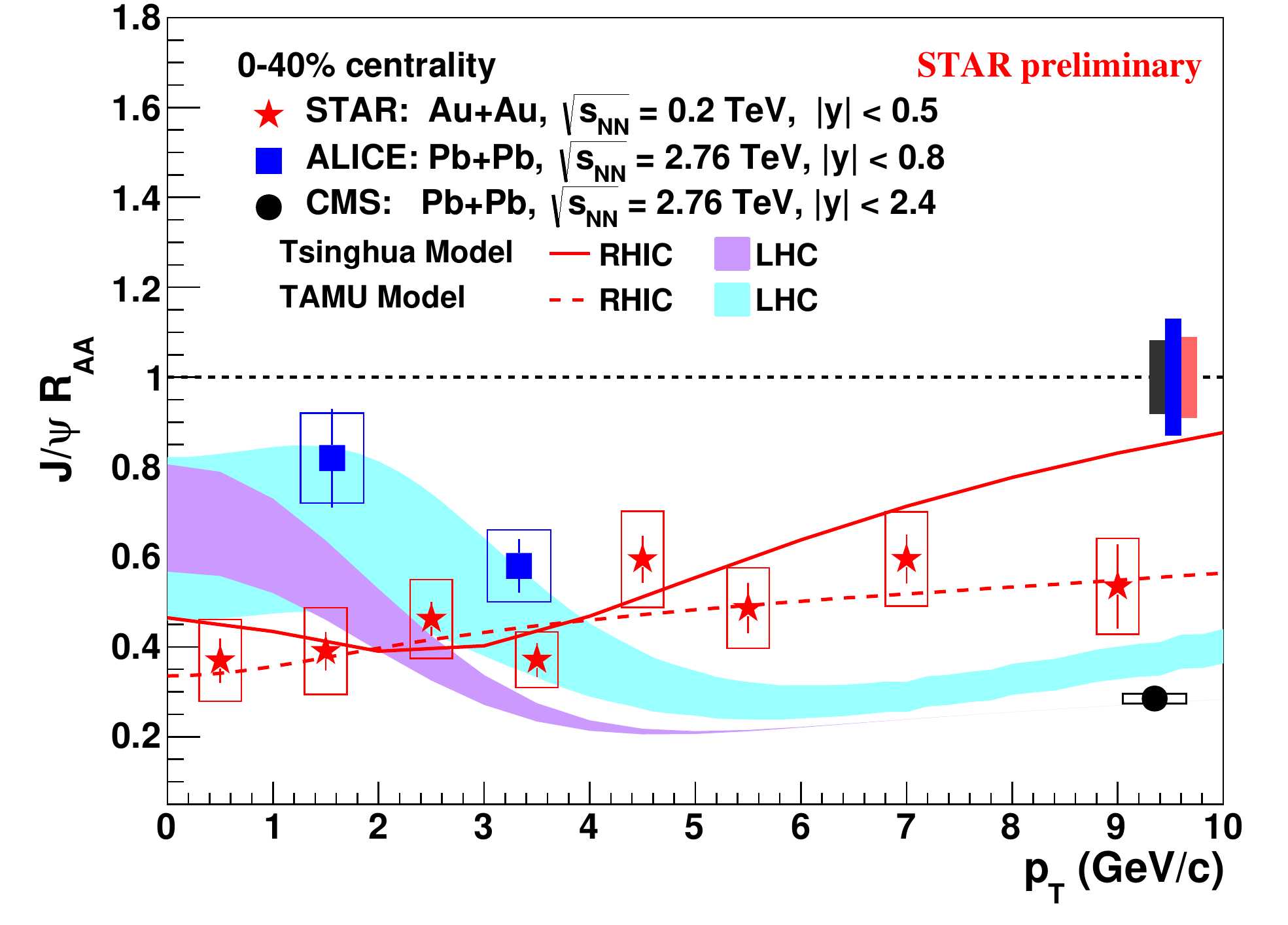}
\caption{\label{fig:JpsiRaaVsPt}
$R_{\rm AA}$ as a function of $p_T$ at the RHIC (red star) 
and the LHC (blue square, black circle).
The lines and bands indicate transport model calculations for RHIC and LHC energies.	
}
\end{minipage}\hspace{1.5pc}%
\begin{minipage}{17pc}
\center
\includegraphics[width=15pc]{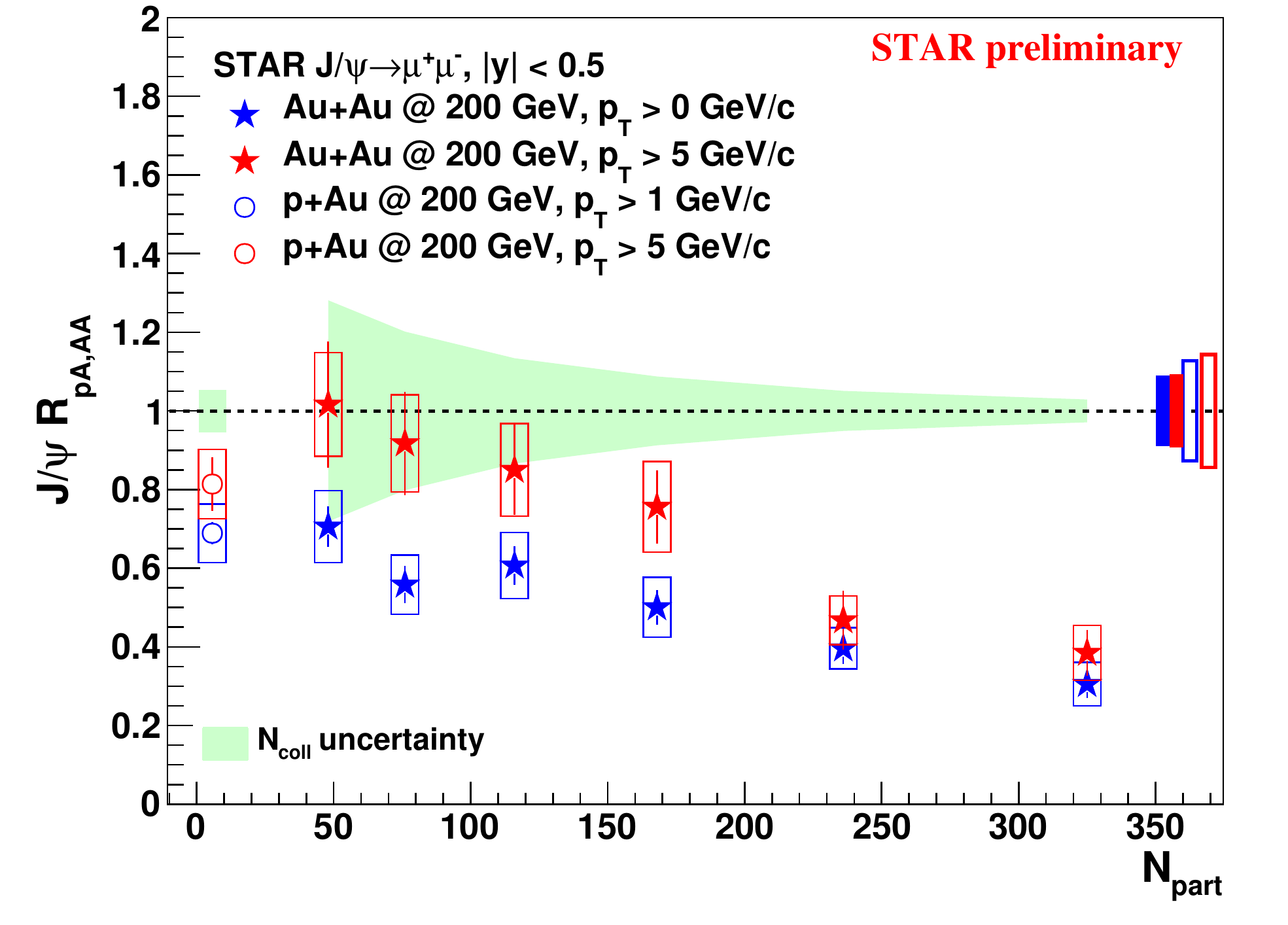}
\caption{\label{fig:Run14RAARun15RpAu} 
$R_{\rm AA}$ (solid star) and $R_{\rm pAu}$ (open circle) for $p_T>0(1)$~GeV/$c$ and $p_T>5$~GeV/$c$ as a function of $N_{\rm part}$.
}
\end{minipage} 
\end{figure}

Shown in Fig.~\ref{fig:JpsiRaaVsPt} is the nuclear modification factor $R_{\rm AA}$ of inclusive $J/\psi$
in 0-40\% central Au$+$Au collisions compared with LHC results~\cite{Abelev:2013ila,Chatrchyan:2012np}.
The strong suppression at RHIC at high $p_T$ indicates significant $J/\psi$ dissociation.
The hint of the increasing $R_{\rm AA}$ with increasing $p_T$ can be explained by the formation-time effect and the feed-down contribution from $B$ hadron decays~\cite{Zhao:2010nk}.
The stronger suppression of $J/\psi$ at RHIC at low $p_T$ can be explained by less regeneration contribution due to smaller charm production cross section,
while the smaller suppression of $J/\psi$ at RHIC at high $p_T$ could arise from a smaller dissociation rate due to the lower temperature of the medium.
The $R_{\rm AA}$ as a function of the number of participant nucleons ($N_{\rm part}$) for $p_T>0$~GeV/$c$ and $p_T>5$~GeV/$c$ are compared 
with the $R_{\rm pAu}$ in Fig.~\ref{fig:Run14RAARun15RpAu}.
The nuclear modification factors in the most peripheral Au+Au collisions are consistent with those measured in p+Au collisions.

\begin{figure}[h]
\begin{minipage}{17pc}
\center
\includegraphics[width=15pc]{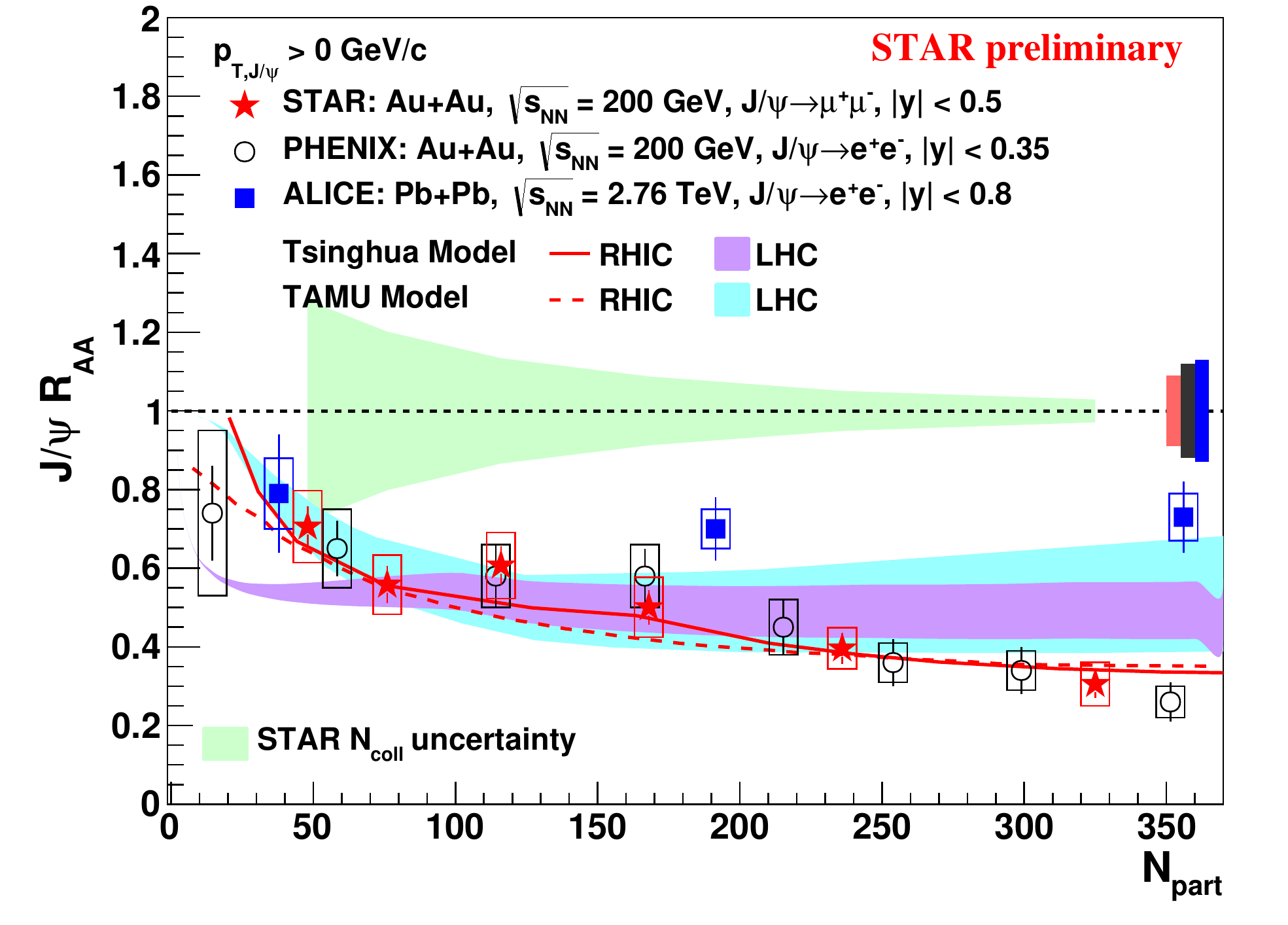}
\caption{\label{fig:JpsiRaaNpartLowPt} Nuclear modification factor $R_{\rm AA}$ for $p_T>0$~GeV/$c$ as a function of $N_{\rm part}$.}
\end{minipage}\hspace{1.5pc}%
\begin{minipage}{17pc}
\center
\includegraphics[width=15pc]{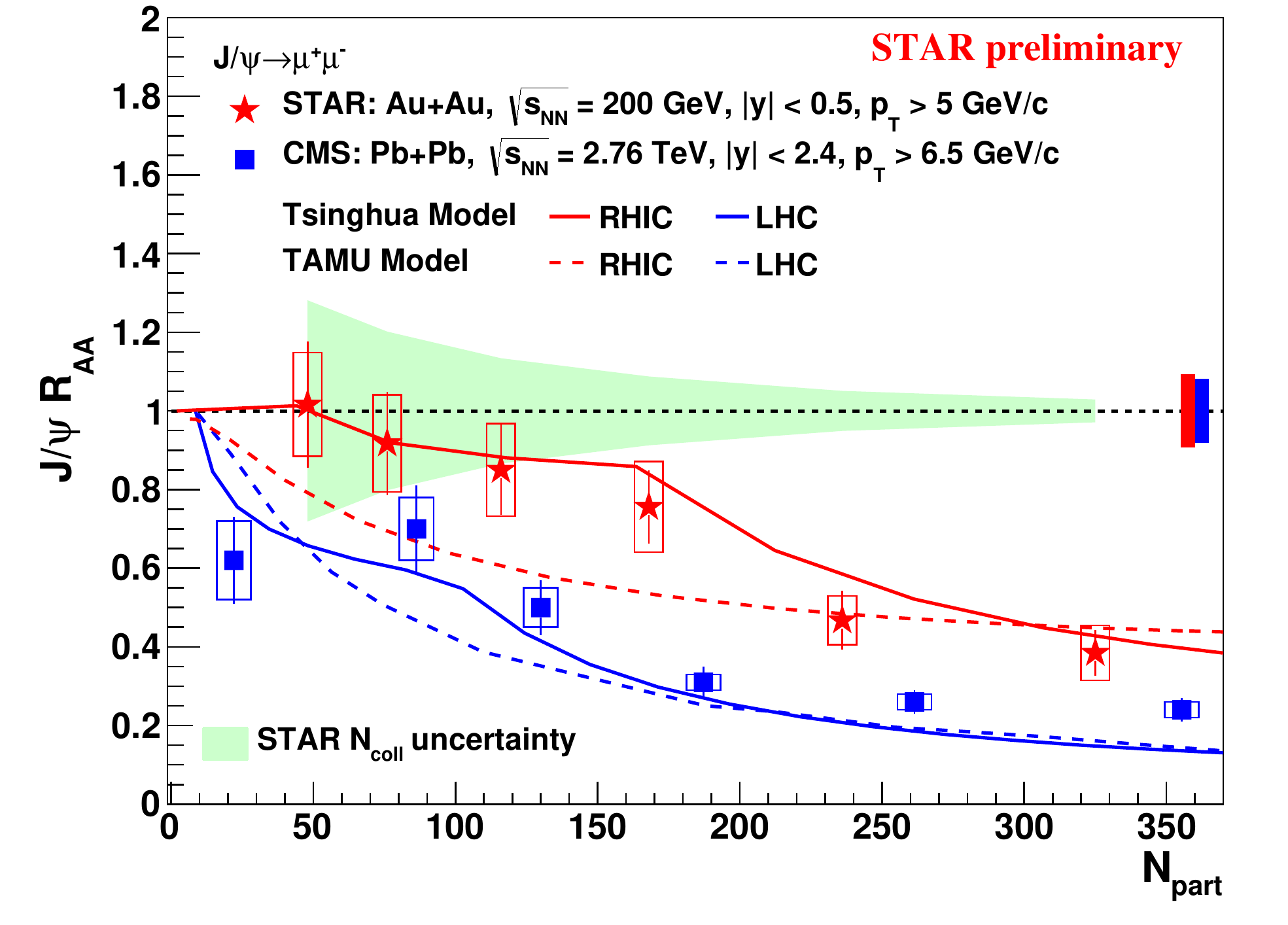}
\caption{\label{fig:JpsiRaaNpartHighPt} Nuclear modification factor $R_{\rm AA}$ for $p_T>5$~GeV/$c$ as a function of $N_{\rm part}$.}
\end{minipage} 
\end{figure}

Transport models from Tsinghua~\cite{Liu:2009nb, Zhou:2014kka} and Texas A\&M University (TAMU) ~\cite{Zhao:2010nk, Zhao:2011cv} groups,
including dissociation and regeneration contributions, can qualitatively describe the $p_T$ dependence of the RHIC and the LHC data as shown in Fig.~\ref{fig:JpsiRaaVsPt}.
Centrality dependences of the $J/\psi$ $R_{\rm AA}$ at the RHIC~\cite{Adare:2006ns} and the LHC are shown in Fig.~\ref{fig:JpsiRaaNpartLowPt} for $p_T>0$~GeV/$c$
and in Fig.~\ref{fig:JpsiRaaNpartHighPt} for $p_T>5$~GeV/$c$.
For $p_T>0$~GeV/$c$, both models can describe the centrality dependence at the RHIC, but tend to overestimate the suppression at the LHC.
For $p_T>5$~GeV/$c$, there is tension among models and data.
The discontinuities seen in the $R_{\rm AA}$ as a function of $N_{\rm part}$ from the Tsinghua model calculation
can be attributed to the complete dissociation of $J/\psi$ when the 
medium temperature exceeds the dissociation temperature.

\section{Summary}
In summary, we presented the first charmonium measurements in the di-muon decay channel at mid-rapidity at the RHIC.
In p$+$p collisions at $\sqrt{s}=200$~GeV, inclusive $J/\psi$ production cross section can be described by CGC+NRQCD and NLO NRQCD model calculations
for prompt $J/\psi$ at low and high $p_T$ ranges, respectively. While the ICEM calculation for direct $J/\psi$ can describe the data for $p_T<3$~GeV/$c$,
it generally underestimates the yield at higher $p_T$. 
In p$+$Au collisions at $\sqrt{s_{_{\rm NN}}}=200$~GeV, we observe
(i) inclusive $J/\psi$ $R_{\rm pAu}$ is consistent with $R_{\rm dAu}$ suggesting similar cold nuclear matter effects in p+Au and d+Au collisions;
(ii) calculations incorporating the nuclear PDF and nuclear absorption effects can well describe $R_{\rm pAu}$;
and
(iii) the double ratio of inclusive $J/\psi$ and $\psi(2S)$ production rates between p$+$p and p$+$Au collisions is 1.37~$\pm$~0.42~$\pm$~0.19.
In Au$+$Au collisions at $\sqrt{s_{_{\rm NN}}}=200$~GeV, we observe
(i) significant $J/\psi$ suppression in central collisions at high $p_T$ indicating dissociation;
(ii) the $J/\psi$ $R_{\rm AA}$ can be qualitatively described by transport models including dissociation and regeneration;
and
(iii) the $R_{\rm AA}$ in the most peripheral collisions is consistent with the $R_{\rm pAu}$.
These measurements in Au$+$Au collisions will gain additional statistical precision by combining with the similar amount of data recorded in the RHIC 2016 run. 





\bibliographystyle{elsarticle-num}
\bibliography{QM17_CharmoniumProceedings_preprint}

\begin{thebibliography}{10}
\expandafter\ifx\csname url\endcsname\relax
  \def\url#1{\texttt{#1}}\fi
\expandafter\ifx\csname urlprefix\endcsname\relax\def\urlprefix{URL }\fi
\expandafter\ifx\csname href\endcsname\relax
  \def\href#1#2{#2} \def\path#1{#1}\fi

\bibitem{Matsui:1986dk}
T.~Matsui, H.~Satz, Phys. Lett. B178 (1986) 416--422.

\bibitem{Ma:2014mri}
Y.-Q. Ma, R.~Venugopalan, Phys. Rev. Lett. 113~(19) (2014) 192301.

\bibitem{Shao:2014yta}
H.-S. Shao, H.~Han, Y.-Q. Ma, C.~Meng, Y.-J. Zhang, K.-T. Chao, JHEP 05 (2015)
  103.

\bibitem{Ma:2016exq}
Y.-Q. Ma, R.~Vogt, Phys. Rev. D94~(11) (2016) 114029.

\bibitem{Adare:2012qf}
A.~Adare, et~al., Phys. Rev. C87~(3) (2013) 034904.

\bibitem{Lansberg:2016deg}
J.-P. Lansberg, H.-S. Shao, Eur. Phys. J. C77~(1) (2017) 1.

\bibitem{Shao:2012iz}
H.-S. Shao, Comput. Phys. Commun. 184 (2013) 2562--2570.

\bibitem{Shao:2015vga}
H.-S. Shao, Comput. Phys. Commun. 198 (2016) 238--259.

\bibitem{Shao:2016private}
H.-S. Shao, R.~Vogt, {Private Communication (2017)}.

\bibitem{Ferreiro:2012sy}
E.~G. Ferreiro, F.~Fleuret, J.~P. Lansberg, N.~Matagne, A.~Rakotozafindrabe,
  Few Body Syst. 53 (2012) 27--36.

\bibitem{Abt:2006va}
I.~Abt, et~al., Eur. Phys. J. C49 (2007) 545--558.

\bibitem{Adare:2011vq}
A.~Adare, et~al., Phys. Rev. D85 (2012) 092004.

\bibitem{Adare:2016psx}
A.~Adare, et~al., Phys. Rev. C95~(3) (2017) 034904.

\bibitem{Abe:1997jz}
F.~Abe, et~al., Phys. Rev. Lett. 79 (1997) 572--577.

\bibitem{Adare:2013ezl}
A.~Adare, et~al., Phys. Rev. Lett. 111~(20) (2013) 202301.

\bibitem{Ferreiro:2016private}
E.~G. Ferreiro, {Private Communication (2016)}.

\bibitem{Ferreiro:2014bia}
E.~G. Ferreiro, Phys. Lett. B749 (2015) 98--103.

\bibitem{Abelev:2013ila}
B.~B. Abelev, et~al., Phys. Lett. B734 (2014) 314--327.

\bibitem{Chatrchyan:2012np}
S.~Chatrchyan, et~al., JHEP 05 (2012) 063.

\bibitem{Zhao:2010nk}
X.~Zhao, R.~Rapp, Phys. Rev. C82 (2010) 064905.

\bibitem{Liu:2009nb}
Y.-p. Liu, Z.~Qu, N.~Xu, P.-f. Zhuang, Phys. Lett. B678 (2009) 72--76.

\bibitem{Zhou:2014kka}
K.~Zhou, N.~Xu, Z.~Xu, P.~Zhuang, Phys. Rev. C89~(5) (2014) 054911.

\bibitem{Zhao:2011cv}
X.~Zhao, R.~Rapp, Nucl. Phys. A859 (2011) 114--125.

\bibitem{Adare:2006ns}
A.~Adare, et~al., Phys. Rev. Lett. 98 (2007) 232301.

\end{thebibliography}







\end{document}